\documentclass[prl,twocolumn,superscriptaddress,unsortedaddress,showpacs]{revtex4}

\usepackage{graphicx}
\usepackage{amsmath}

\begin{document}

\title{Spatial Coherence Resonance near Pattern-Forming Instabilities}

\author{O. Carrillo}

\affiliation{Departament d'Estructura i Constituents de la Mat\`eria,
Universitat de Barcelona, Diagonal 647, E--08028 Barcelona, Spain}

\author{Miguel A. Santos}

\affiliation{Departament d'Enginyeria Mec\`anica, Universitat Rovira i
Virgili, Carretera Salou s/n, 43006 Tarragona, Spain}

\author{J. Garc\'{\i}a--Ojalvo}

\email{jordi.g.ojalvo@upc.es}

\affiliation{Center for Applied Mathematics, 657 Rhodes Hall,
Cornell University, Ithaca, NY 14853}

\affiliation{Departament de F\'{\i}sica i Enginyeria Nuclear,
Universitat Polit\`ecnica de Catalunya,
Colom 11, E--08222 Terrassa, Spain}

\author{J.M. Sancho}

\affiliation{Departament d'Estructura i Constituents de la Mat\`eria,
Universitat de Barcelona, Diagonal 647, E--08028 Barcelona, Spain}

\date{\today}

\begin{abstract}
The analogue of temporal coherence resonance for spatial degrees of freedom is
reported. Specifically, we show that
spatiotemporal noise is able to optimally extract an intrinsic spatial
scale in nonlinear media close to (but before) a pattern-forming instability.
This effect is observed in a model of pattern-forming chemical reaction
and in the Swift-Hohenberg model of fluid convection.
In the latter case, the phenomenon is described analytically via
an approximate approach.
\end{abstract}

 
\maketitle

The ability of noise to induce temporal coherence in nonlinear systems is
a well documented fact. In stochastic resonance (SR), for instance, random
fluctuations enhance the response to weak periodic driving, as observed in
many different physical, chemical and biological scenarios \cite{wiesenfeld}.
From a more intriguing perspective, even systems showing no explicit time
scale exhibit an enhancement of temporal coherence due to noise
\cite{gang,rappel}. This {\em autonomous} stochastic resonance, where noise
extracts a hidden, intrinsic time scale of the system's dynamics,
has been termed coherence resonance (CR) \cite{pikovsky}, and has been
predicted theoretically in a wide variety of models (see \cite{revexc}
for a review) and observed experimentally in fields as diverse as
plasma physics \cite{i}, laser dynamics \cite{jorge}, electronics
\cite{postnov}, atom trapping \cite{wilkowski} and neuroscience
\cite{gu}.

Parallel to the advances in the fields of stochastic and coherence
resonance, recent years have witnessed an increasing interest in the
dynamics of systems with spatial degrees of freedom.
Much work has been devoted to analyze the enhancement of spatiotemporal
order by noise \cite{nises,alonso}, and the phenomena of
standard temporal SR and CR have been studied in detail in arrays of
dynamical elements \cite{sr-ext,cr-ext}, where it has been shown
that coupling enhances the noise-induced temporal coherence. But so far,
up to our knowledge,
no analogues of CR (or SR, for that matter) for spatial degrees of
freedom exist. Namely, we ask ourselves if noise is able to extract
and optimize an intrinsic {\em spatial} scale from a deterministically
homogeneous extended medium. In this Letter we show that such {\em
spatial coherence resonance} (SCR) exists close to pattern-forming
instabilities, mimicking one of the mechanisms of standard (temporal) CR.

There are different mechanisms giving rise to temporal CR. In the most
extended situation, CR arises as the result of a matching between the
time scale associated to noise-induced escape from a stable steady
state and the characteristic time scale of the phase-space trajectories
outside from the immediate vicinity of the fixed point (such as 
what happens in excitable media \cite{pikovsky,benji} or close
to a saddle-node bifurcation in phase models \cite{rappel}).
A second, less studied mechanism takes place near the onset of
dynamical instabilities giving rise to periodic behavior, where
noise excites precursors of the bifurcation in the form of stochastic
limit cycles with a well-defined peak in the power spectrum, even before
the bifurcation really happens \cite{kurt85,omberg}.
Such {\em noisy precursors} exhibit an optimal coherence at an
intermediate noise strength, in a way typical of coherence resonance
\cite{shura}. This effect has been confirmed experimentally \cite{kiss}.

Analogously to the temporal case, noisy precursors of pattern-forming
bifurcations have been observed, in the form of noise-induced {\em spatial}
structures, in electroconvection \cite{rehberg}, Rayleigh-B\'enard
convection \cite{wu} and optics \cite{miguel}, for instance. In this
case, the spatial power spectrum (now structure function) of
the system exhibits a well defined peak due to noise, even below
the bifurcation point. In what follows, we show that noise controls
the shape of this spectral peak in a manner characteristic of coherence
resonance, but with the role of temporal frequency being played by spatial
wave number. This allows us to define a pure spatial analogue of
temporal coherence resonance.
The generality of the phenomenon is illustrated through its occurrence
in two different pattern-forming systems, namely
an activator-inhibitor model of chemical pattern formation and
the Swift-Hohenberg model of fluid convection. In the latter case the effect 
is interpreted analytically.

We first illustrate the phenomenon of SCR in chemical pattern formation,
in particular in the case of the chlorine dioxide-iodine-malonic acid (CDIMA)
reaction, which is well known to exhibit both temporal and spatial dynamics
\cite{Eps98}. Experimental and theoretical studies of this reaction have
concluded that the malonic acid serves only as a source of iodide, and
thus the system can be described by just two variables, in the way of
an activator-inhibitor model: \cite{Eps98,ancherlengues}
\begin{eqnarray}
\dot{u}&=&{\bf\nabla}^2u\,+a\,-\,c\,u\,-4\,\frac{u\,v}{(1\,+\,u^2)}\,-\,\phi \nonumber \\
\dot{v}&=&b\,\left\{{\bf\nabla}^2v\,+\,c\,u\,-\frac{u\,v}{(1\,+\,u^2)}\,+\,
\phi\right\}\,,
\label{CDIMA}
\end{eqnarray}
where $u$ and $v$ play the role of activator and inhibitor fields,
corresponding to the concentrations of iodine and chlorine dioxide ions,
respectively (all variables and parameters are in dimensionless units). 
This model describes the photosensitivity of the CDIMA
reaction, with $\phi$ accounting for the illumination level, which is
an externally controlled parameter. Additionally, for large differences in
the diffusion coefficients of the two species this system presents a
Turing instability. In particular, the model has an homogenous stable
state given by $u_h=(a-5\phi)/(5c)$ and $v_h=a(1+u_h^2)/(5u_h)$, which
becomes unstable versus spatial perturbations of nonzero wavenumber
for $\phi<\phi_c$. For $a=16$, $b=301$, $c=0.6$ and
$d=1.07$, the critical illumination is $\phi_c=2.3$ \cite{ancherlengues}.

\begin{figure}[htb]
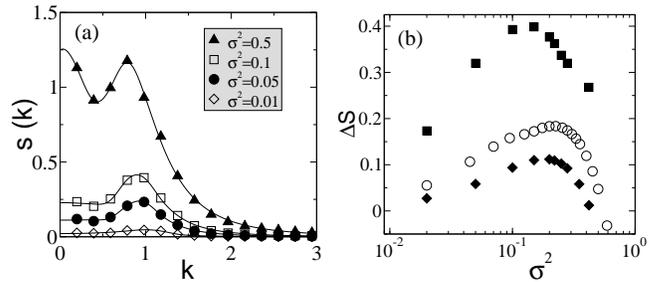

\centerline{
\includegraphics[width=0.23\textwidth,clip]{fig1a.eps}
\hskip1mm
\includegraphics[width=0.23\textwidth,clip]{fig1b.eps}}
\caption{(a) Spherical average of the structure function for increasing noise
intensities. Average illumination is $\phi=2.35$.
(b) $\Delta S$ versus noise intensity for different values of the
average illumination: $\phi=2.32$ (squares), $\phi=2.35$ (circles), and
$\phi=2.37$ (diamonds). The simulations have been performed on a
square lattice of 64$\times$64 cells with mesh size $\Delta x=0.5$.}
\label{SrPeak}
\end{figure}

This chemical system allows for an experimental study of stochastic effects,
by introducing noise in the illumination profile. In this way, for instance,
it has been shown experimentally that a quenched spatial noise sustains
Turing patterns in this system well beyond the deterministic bifurcation
threshold \cite{ancherlengues}. We now consider a spatiotemporally stochastic
illumination profile, $\phi\to\phi\,+\,\xi$, where $\xi(\vec{x},t)$ is a
Gaussian noise with zero mean, white in time and with spatial autocorrelation
given by
\begin{equation}
\langle\xi(\vec{x},t)\xi(\vec{x}',t')\rangle\,=\,2\,\sigma^2\,
C(\vec{x}-\vec{x}')\,\delta(t-t')
.\label{ncorr}
\end{equation}
In order to examine the spatial response of the system in the presence
of noise, we evaluate the structure function of the activator variable,
defined as
\begin{equation}
S(\vec{k},t)\,=\,\frac{\langle \widehat{u}(\vec{k},t)\,
\widehat{u}(-\vec{k},t)\rangle}{V}\,,
\label{Str}
\end{equation}
where $\widehat{u}(\vec{k},t)$ is the spatial Fourier transform
of the activator field, $V$ is the d-dimensional volume of the system,
and $\langle\cdots\rangle$ represents an ensemble average over noise
realizations. Taking into account the spherical symmetry of this
system, we compute the
spherical average of the structure function as $s(k)\,=\,
\int_{\Omega_k}d\Omega_k\,S(\vec{k},t)$, where $k=\left|\vec{k}\right|$
and $\Omega_k$ is a hyperspherical shell of radius $k$.
The result is shown in Fig.~\ref{SrPeak}(a) for increasing noise intensities
\cite{dclevel}, evaluated for a two-dimensional
system and in the case of spatially white noise, i.e.
$C(\vec{x}_i-\vec{x}_j)\,=\,\delta_{ij}/\Delta x$.
The average illumination $\phi$ to which the system is subjected is
above threshold, thus for vanishing or very weak noises the system is
homogeneous and the structure function does not show a well-defined peak
as $k$ increases.
However, increasing the noise intensity induces a peak in the structure
function at $k=k_{\rm max}$, which represents a spatial noisy precursor
of the pattern-forming
bifurcation that occurs at $\phi=\phi_c$: fluctuations are capable to
anticipate the instability, making the structure manifest even before
the threshold has been crossed.
If fluctuations become too large, disorder inevitably comes into
play, and the peak in the structure function becomes less well defined
[see triangles in Fig.~\ref{SrPeak}(a)]. It is then clear that an
optimal noise intensity exists at which the structure function peak is
best resolved from the background fluctuations. In order to quantify the
efficiency of the noise, an adequate measure must be used. There are
different ways of quantifying such a response depending on the peculiarities
of the problem under study. Here, we estimate the spatial coherence of the
system through the quantity $\Delta S \equiv s(k_{\rm max})-s(0)$, taking
into account that $s(0)$ measures the level of fluctuations existing in
the system. 
\begin{figure}[htb]
\centerline{
\includegraphics[width=0.15\textwidth,clip]{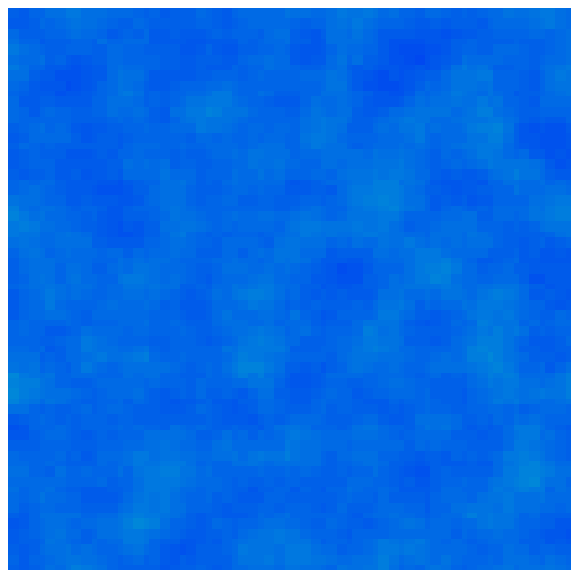}
\includegraphics[width=0.15\textwidth,clip]{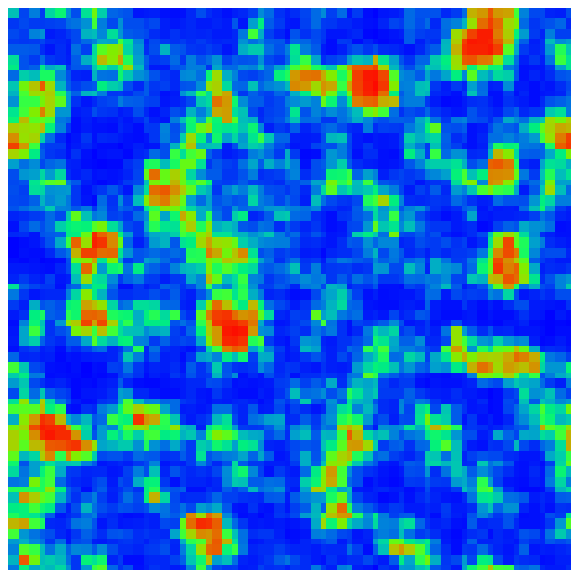}
\includegraphics[width=0.15\textwidth,clip]{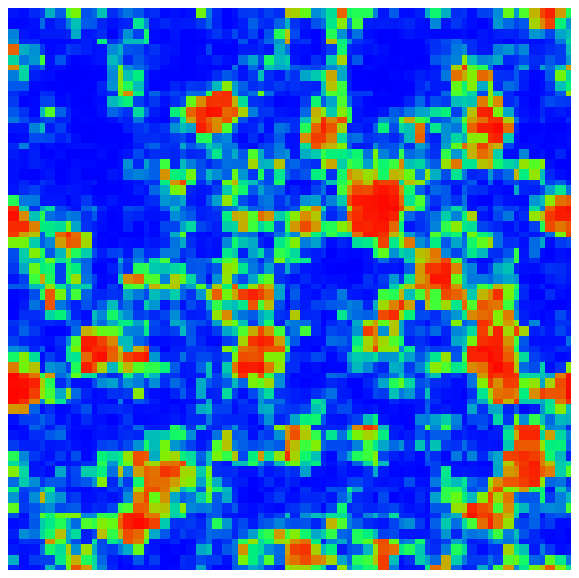}
}
\vspace{0.2cm}
\caption{Spatial profiles of the activator field for increasing noise
strength and $\phi=2.35$, with color mapped nonlinearly (with a $\tanh$
function) and the same contrast in all figures for comparison.
From left to right, $\sigma^2=0.001, 0.2, 0.6$. The
lattice parameters are those of Fig.~\protect\ref{SrPeak}.}
\label{SNDp}
\end{figure}
Figure~\ref{SrPeak}(b) shows how the SND varies
with noise intensity $\sigma^2$ for different illumination levels. 
It can be seen that there is an optimal noise strength $\sigma^2_{\rm opt}$ for
which the structure function peak is resolved best. This is the signature
of a spatial coherence resonance driven by the external stochastic forcing,
which extracts the intrinsic spatial distribution of the system. As shown in
Fig.~\ref{SrPeak}(b), the closer the system is to the bifurcation point,
the more pronounced the resonance is, and the position of the maximum
$\sigma^2_{\rm opt}$ shifts to smaller values.

Figure~\ref{SNDp} shows several patterns produced by model (\ref{CDIMA})
for different values of the noise strength at $\phi=2.35$. While the
system is quasi-homogenous for small noise (left plot),
it exhibits a rather self-organized configuration for
$\sigma^2\sim\sigma^2_{\rm opt}$ (middle plot), and becomes gradually
more disordered for large noise strengths (right plot).
Although the effect may not appear very clear in the spatial profiles,
the relevant fact is the appearance and enhancement of the peak in the
structure function,
since this is a measurable effect through, for instance, scattering
experiments.

\begin{figure}[htb]
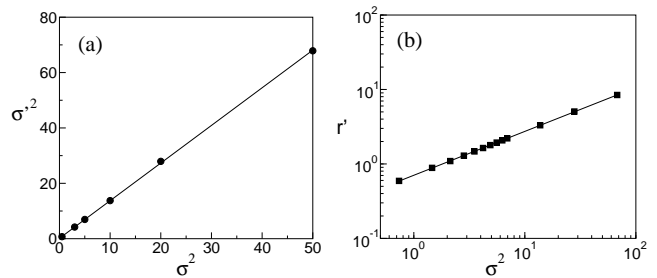

\centerline{
\includegraphics[width=0.23\textwidth,clip]{fig3a.eps}
\hskip1mm
\includegraphics[width=0.23\textwidth,clip]{fig3b.eps}}
\caption{Effective structure-function parameters for model
(\protect\ref{SHeq}), for $r=0.05$ and $k_0=0.6$.
In (a), the solid line corresponds to the linear relation
$\sigma'^2 = 1.365 \sigma^2$, and in (b) to
$r'= 0.05+a(\sigma^2)^{\lambda}$, with
$a \simeq 0.704$ and $\lambda \simeq 0.6$.}
\label{fig:shfit}
\end{figure}

We may illustrate analytically this phenomenon in the case of a simpler
pattern-forming model, namely the Swift-Hohenberg
equation, which describes the onset of hydrodynamic convection in
Rayleigh-B\'enard cells \cite{cross}:
\begin{equation}
\frac{\partial \phi(\vec x,t)}{\partial t} = -r \phi - \phi^{3} -
\left(\nabla^{2} +
k_{0}^{2}\right)^{2} \phi + \xi(\vec x,t)\,,
\label{SHeq}
\end{equation}
where the control parameter $r$ is proportional to the temperature gradient
driving the fluid, and the additive noise $\xi(\vec x,t)$ is again considered
Gaussian and white in both time and space. As in the CDIMA model,
in the absence of noise and for $r>0$,
this system has a homogeneous steady state $\phi=0$,  
representing conduction, 
which upon decrease of $r$ becomes
unstable at $r=r_c=0$ versus static perturbations of nonzero wavenumber $k=k_0$,
which represents convection \cite{cross}. In the presence of a small amount of noise, the structure
function can be calculated by linearizing Eq.~(\ref{SHeq}) around $\phi=0$,
and transforming the system to Fourier space \cite{nises}. The result is:
\begin{equation}
S_{st}(k) = \frac{\sigma^2}{(k^{2}-k_{0}^{2})^{2}+r}
\label{struct}
\end{equation}
Since we want to operate in regimes where, even though the system is
deterministically homogeneous, noise can be large, we generalize the
previous expression (\ref{struct}) by allowing the parameters $\sigma$
and $r$ to depend on noise; we will call them $\sigma'(\sigma)$ and
$r'(\sigma)$ in what
follows. In order to estimate how these renormalized parameters depend on
the noise intensity, we rederive expression (\ref{struct}) without
neglecting completely nonlinear terms. To do that, we estimate the cubic
term in (\ref{SHeq}) by applying the Gaussian approximation \cite{langer}
$\phi(x,t)^3 \sim 3 \phi(x,t) \langle \phi^2 \rangle_{st}$. This renders
back the equation linear, so that the structure function obeys again
expression (\ref{struct}) with $r$ being replaced by
$r'\equiv r+3\langle \phi^2 \rangle_{st}$. We can now calculate the value of
the average squared field $\langle \phi^2 \rangle_{st}$ from the
self-consistency relation
\begin{eqnarray}
\langle \phi^2 \rangle_{st} &=& \frac{1}{(2 \pi)^2}\int^{\infty}_0 S_{st}(k) 
2\pi k\, dk
\nonumber\\
&=&\frac{\pi \sigma^2}{(2 \pi)^2 \, \sqrt{r'}}\left[ \frac{\pi}{2} + 
tg^{-1}\left(\frac{k_0^2}{\sqrt{r'}}\right)\right] \sim
\frac{\sigma^2}{8 \sqrt{r'}}.
\label{eq:sk}
\end{eqnarray}
Since we will be operating close to threshold, $r$ will be small
and $r' \sim 3\langle \phi^2 \rangle_{st}$, which leads from (\ref{eq:sk})
to $\langle \phi^2 \rangle_{st} \sim  (\sigma^2)^{2/3}$. Therefore,
from the definition of $r'$ above we finally obtain that this
effective parameter also scales with noise intensity as
$r' \sim  (\sigma^2)^{2/3}$. In order to verify this result, we perform
numerical simulations of model (\ref{SHeq}) in the homogeneous regime
for different noise intensities, and fit the resulting structure
functions with expression (\ref{struct}), with effective parameters
$\sigma'$ and $r'$ which are allowed to depend on noise. The results
of this analysis are shown in Fig.~\ref{fig:shfit}, and confirm that
$\sigma'$ scales linearly with noise intensity $\sigma$ (as expected,
since nonlinear corrections did not affect this parameter in the
analysis made above), whereas $r'$ scales with noise intensity with
an exponent fairly close to 2/3. 
 
\begin{figure}[htb]
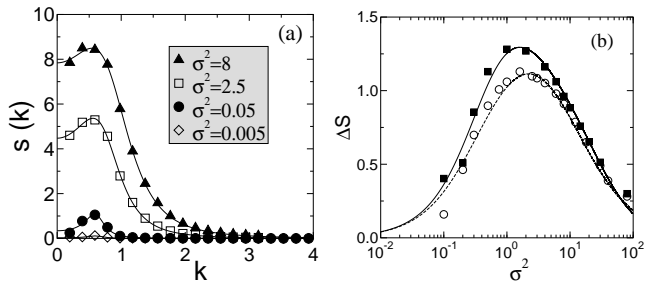

\centerline{
\includegraphics[width=0.23\textwidth,clip]{fig4a.eps}
\hskip1mm
\includegraphics[width=0.23\textwidth,clip]{fig4b.eps}
}
\caption{(a) Structure function exhibited by the Swift-Hohenberg model
(\protect\ref{SHeq}) for $r=0.005$, $k_0=0.6$ and four different noise
intensities, listed in the legend.
(b) $\Delta S$ versus noise intensity for two different values of
the control parameter $r$. Squares correspond to $r=0.005$ and
circles to $r=0.05$. In the two plots, lines are the fitted
function (\protect\ref{snd1}) with the renormalized parameters
given in Fig.~\protect\ref{fig:shfit}. The simulations have been
performed again in a square lattice of 64$\times$64 cells with
spacing $\Delta x=0.5$.}
\label{fig:shsk}
\end{figure}

The numerical results of Fig.~\ref{fig:shsk}(a) have been fitted with the analytic
approximation given by Eq.~(\ref{struct})) with renormalized parameters,
leading to a quite good agreement. This is shown in
Fig.~\ref{fig:shsk}(a) for four different noise intensities. Similarly
to what occurs in the chemical model, we can see how noise excites
a peak in the structure function (in this case at $k=k_0$) even within the homogeneous
regime. This peaks increases in size as
noise intensity increases, both in an absolute way and relative to the
value of the structure function at $k=0$. However, for large noise
intensities disorder kicks in again, and the peak becomes less pronounced.
This effect can be seen quantitatively, in terms of the dependence of
$\Delta S$ on $\sigma^2$, for two different distances to threshold
in Fig.~\ref{fig:shsk}(b), where the numerical results are again
compared to the theoretical prediction coming from the renormalized
version of (\ref{struct}), which in this case gives
\begin{equation}
\Delta S(\sigma^2) = \frac{k_{0}^4 \sigma'^2}{r'(k_{0}^4+r')}
\label{snd1}
\end{equation}
The agreement between the numerical results and the theoretical prediction
is rather satisfactory. In all cases, a clear enhancement of spatial
coherence (in terms of spatial structures with wavenumber $k_{\rm max}=k_0$)
is evident.

In conclusion, we have demonstrated that spatiotemporal noise is able
to extract and enhance spatial coherence from deterministically
homogeneous nonlinear media. This effect constitutes a pure spatial
analogue of temporal coherence resonance, and relies on the generation
of spatial noisy precursors of pattern-forming instabilities.
This phenomenon is different from what happens in noise-induced
pattern formation \cite{nipf}, where special types of multiplicative
noise induce (or displace) the pattern-forming transition, destabilizing
the homogeneous state. Here that state is still stable, and the noise
can be simply additive. Additionally, our
results lead us to expect that spatial analogues of both stochastic
resonance and other mechanisms of coherence resonance (such as the one
that takes place in excitable systems \cite{pikovsky}) could
be found. The latter perspective, given the ubiquity of excitable systems
in all areas of science \cite{revexc}, is in our opinion very attractive.
Excitable neural tissue, for instance, combines the features of being
highly noisy and intrinsically spatially extended \cite{kleinfeld}.
Elucidating whether noise is able to enhance the {\em spatial}
coherence of the system in this context would be extremely interesting.

We thank L. Schimansky-Geier for useful comments.
This research was supported by the Ministerio de Ciencia y Tecnolog\'{\i}a
(Spain) and FEDER under projects BFM2000-0624, BFM2001-2159, and
BFM2002-04369. J.G.O. is partially supported by the NSF IGERT Program
of Nonlinear Sciences (Cornell).


\begin{thebibliography}{99}
\bibitem{dclevel} The homogenous component of the spatial profile in this
system gives rise to a spike of $s(k)$ at $k=0$, which has been extracted
in the present results. The final value of $s(0)$ used here is then
obtained by extrapolating to $k=0$ a fitting function defined for $k>0$.
\end{thebibliography}
\end{document}